\documentclass[aps,prb,twocolumn,showpacs,amsfonts,amssymb,amsmath,superscriptaddress]{revtex4}
\usepackage{bm}
\usepackage{graphicx}
\usepackage{longtable}

\begin{document}

\title{From the bulk to monatomic wires: An {\it ab-initio} study of
magnetism in Co systems with various dimensionality}

\date{\today}

\author{Matej Komelj}
\affiliation{Brookhaven National Laboratory, Center for  Data Intensive
  Computing, Upton, 11973 New York (USA)}
\affiliation{Jo\v zef Stefan Institute, Jamova 39, SI-1000 Ljubljana,
  Slovenia}
\author{Claude Ederer}
\affiliation{Max-Planck Institut f\"ur Metallforschung, Heisenbergstr.~1,
  D-70569 Stuttgart, Germany}
\author{James W. Davenport}
\affiliation{Brookhaven National Laboratory, Center for  Data Intensive
  Computing, Upton, 11973 New York (USA)}
\author{Manfred F\"ahnle}
\email{faehn@physix.mpi-stuttgart.mpg.de}
\homepage{http://physix.mpi-stuttgart.mpg.de/schuetz/elth/electronth.html}
\affiliation{Max-Planck Institut f\"ur Metallforschung, Heisenbergstr.~1,
  D-70569 Stuttgart, Germany}

\begin{abstract}
A systematic {\it ab-initio} study within the framework of the
local-spin-density approximation including spin-orbit coupling and an
orbital-polarization term is performed for the spin and orbital
moments and for the X-ray magnetic circular dichroism (XMCD) spectra
in hcp Co, in a Pt supported and a free standing Co monolayer, and in
a Pt supported and a free standing monatomic Co wire. When including
the orbital-polarization term, the orbital moments increase
drastically when going to lower dimensionality, and there is an
increasing asymmetry between the $L_2$ and $L_3$ XMCD signal. It is
shown that spin and orbital moments can be obtained with good accuracy
from the XMCD spectra via the sum rules. The $\langle T_z \rangle$
term of the spin sum rule is surprisingly small for the wires, and the
reason for this is discussed.
\end{abstract}


\maketitle

The modern methods to prepare nanostructured systems made it possible
to investigate the influence of dimensionality on the magnetic
properties.  A central question thereby is how the qualitative
behavior will change when going from two-dimensional to
one-dimensional systems because it has been predicted that there is no
long-range magnetic order at finite temperature in infinitely extended
one-dimensional systems with short-range magnetic interactions. In the
past, several experimental investigations of monolayer nanostripes of
Fe on vicinal surfaces of W \cite{Elmers94,Hauschild98} or Cu
\cite{Shen97} have been performed, with a stripe width down to 1-10
nm. Most recently, Gambardella {\it et al.}  \cite{Gambardella00}
succeeded to prepare a high density of parallel atomic chains along steps by
growing Co on a high purity Pt(997) vicinal surface in a narrow
temperature range.  The magnetism of the Co wires was investigated
\cite{Gambardella02} by the X-ray magnetic circular dichroism
(XMCD). Below a blocking temperature a long-range magnetic ordering
owing to the presence of anisotropy barriers was found on the time
scale of the experiment. Applying a simple model of exchange coupled
superparamagnetic clusters \cite{comment1}, the anisotropy energy
could be obtained from the shape of the magnetization curve above the
blocking temperature and it appeared to be much larger than the one
for a Co monolayer on Pt which --- in turn --- is much larger than the
one of hcp Co. Accordingly, a large orbital moment of $0.68\>\mu_{\rm
B}$ per Co atom was found, the highest value ever reported for a 3d
itinerant electron system.
\par
So far, magnetism in quasi-one-dimensional systems was studied mainly
in the insulating material ${\rm CsNiF_3}$ where the magnetic Ni ions
are arranged along linear chains which are well separated from each
other so that the interchain interaction is only $10^{-3}$ or less of
the intrachain interaction. Because of the one-dimensional character
of this spin system and the easy-plane anisotropy, magnetic solitons
play an important role for the dynamical and thermodynamical behavior
which has been investigated by neutron scattering experiments
\cite{Mikeska91}. The discovery of magnetism in one-dimensional
metallic systems opens up the chance to extend the research in many
respects. First, by considering various vicinal surfaces the distance
between the steps and hence the chains can be modified so that the
transition from the well-isolated chains to interacting chains can be
studied. Second, it is possible to grow, e.g., biatomic wires along
the steps and to manipulate the wire length, and to study the
respective influence on magnetic properties. Finally, we expect that
the damping of magnetic excitations is larger for metals than for
insulators and especially large for one-dimensional metals, possibly
leading to peculiar properties of the nonlinear spin excitations
\cite{Holyst86}.
\par
The already existing {\it ab-initio} calculations for monatomic
transition metal wires considered magnetic moments
\cite{Weinert83,Buhlmayer?, Dorantes-Davila98,Zhou99,Spisak02},
exchange couplings \cite{Spisak02} and magnetic anisotropies
\cite{Dorantes-Davila98,Zhou99,Eisenbach02}. It turned out
\cite{Zhou99} that in these one-dimensional systems the orbital
correlations were essential and that the magnetic anisotropy and hence
also the orbital moments could not be calculated reliably by the
local-spin-density approximation (LSDA) in combination with spin-orbit
(SO) coupling. Instead, a term correcting explicitly for the orbital
correlations has to be added.  It is one of the objectives of the
present paper to work out the growing importance of the orbital
correlations with decreasing dimensionality by calculating orbital
(and spin) moments by the LSDA \cite{Perdew92} including SO coupling
(in a self-consistent manner) without and with the
orbital-polarization term \cite{Eriksson90} which takes into account
at least in part the orbital-correlation effects. The calculations are
performed for Co atoms in hcp Co, in a Pt supported and in a free
standing monolayer, and in a Pt supported and in a free standing
monatomic wire.  Another objective is to calculate the respective XMCD
spectra and discuss the influence of dimensionality on the accuracy of
the spin and orbital moments when these are obtained from the XMCD
spectra via the sum rules \cite{Thole92,Carra93}.
\par
To focus on the pure effect of the dimensionality we fix the
nearest-neighbor distance of the atoms for all considered systems to
the one of fcc Pt (2.77\AA). For the monolayers and wires we perform
supercell calculations, i.e., large unit cells which model the
structure are repeated periodically. In the case of the monolayer the
supercell consists of two Pt and one Co \{111\} layers in
the fcc stacking and a vacuum sheet corresponding to two
empty layers. The vicinal Pt(997) surface with the Co wire at the
steps is modeled by the supercell shown in Fig. 1 with two additional
vacuum layers on the top.
\begin{figure}
\includegraphics[scale=2.5]{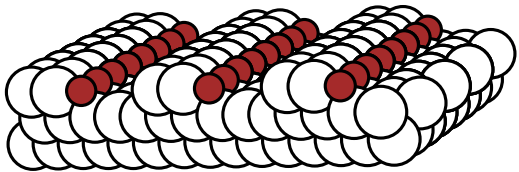}
\caption{The supercell used to model the monatomic Co wire on a
vicinal Pt(997) surface.}
\end{figure}
The supercell of the free standing wire is constructed by removing all
Pt atoms from this supercell. Test calculations have shown that the
results change only very slightly when going to larger supercells. For
the monolayers and the wires we assume perpendicular magnetization,
for hcp Co it is along the c axis.
\par
We use the tight-binding linear-muffin-tin-orbital (LMTO) method in
the atomic sphere approximation \cite{Andersen84/2} in which we have
implemented the SO coupling and the tools for calculating the XMCD
spectra, and the WIEN97 code \cite{Blaha90} which adopts the
full-potential linearized-augmented-plane-wave (FLAPW) method
\cite{Wimmer81} in which the SO coupling and the tools for the
calculation of the magnetooptical effects and XMCD spectra
\cite{Kunes01/1,Kunes01/2} and the orbital-polarization term
\cite{Rodriguez01} have been implemented.  The magnetic orbital
moments $m_l=-\mu_{\rm B}\langle l_z\rangle$ and spin moments
$m_s=-\mu_{\rm B}\langle \sigma_z\rangle$ are calculated directly from
the wave functions as well as via the sum rules \cite{Thole92,Carra93}
from the absorption coefficients $\mu^+(\epsilon)$, $\mu^-(\epsilon)$
and $\mu^0(\epsilon)$ for light with right-circular, left-circular and z-axis
polarization at the $L_2$ and $L_3$ edge according to
\begin{eqnarray}
\label{orbsumrule} \langle l_{z}\rangle & = &
\frac{2I_{\text{m}}N_{\text{h}}}{I_{\text{t}}} \quad , \\
\label{spinsumrule} \langle \sigma_{z}\rangle & = &
\frac{3I_{\text{s}}N_{\text{h}}}{I_{\text{t}}} - 7\langle T_{z}\rangle
\quad , \\ 
I_{\text{m}} & = & \int\limits_{E_{\text{F}}}^{E_{\text{c}}} \left[
(\mu_{\text{c}})_{L_3} + (\mu_{\text{c}})_{L_2} \right] d\epsilon \quad , \\
I_{\text{s}} & = & \int\limits_{E_{\text{F}}}^{E_{\text{c}}} \left[
(\mu_{\text{c}})_{L_3} - 2(\mu_{\text{c}})_{L_2} \right] d\epsilon \quad ,
\\ 
I_{\text{t}} & = & \int\limits_{E_{\text{F}}}^{E_{\text{c}}} \left[
(\mu_{\text{t}})_{L_3} + (\mu_{\text{t}})_{L_2} \right] d\epsilon \quad ,
\end{eqnarray}
with the XMCD signal $\mu_{\text{c}}=\mu^+-\mu^-$ and with $\mu_{\text{t}}=
\mu^++\mu^-+\mu^0$. Here $N_{\text{h}}$ is the number of d holes and
$\langle T_z\rangle$ is the expectation value of the magnetic 
dipolar operator
\begin{equation}
\hat{T}_{z} = \frac{1}{2} [ \bm{\sigma} -
3\hat{\mathbf{r}}(\hat{\mathbf{r}}\cdot\bm{\sigma}) ]_{z} \quad ,
\end{equation}
where $\bm{\sigma}$ denotes the vector of the Pauli matrices. The
quantities $E_{\text{F}}$ and $E_{\text{c}}$ denote the Fermi energy
and a cutoff energy. For details for such type of calculations see
Ref. \onlinecite{Ederer02}. The $\langle T_z\rangle$ term is
negligible for cubic surroundings but it is expected to become more
and more important when reducing the dimensionality of the system.
Since it is very difficult to measure $\langle T_z\rangle $, this term
is often neglected in the spin sum rule when analyzing the
experimental data. One of the objectives of this paper is to asses
this critically.
\par
The results of the LSDA calculations with SO coupling are given in 
the upper part of Table 1. 
\begin{table*}
\begin{ruledtabular}
\begin{tabular}{lrrrrrrrrrrr}

& \multicolumn{2}{c}{hcp} & \multicolumn{2}{c}{monolayer on Pt} &
\multicolumn{2}{c}{free monolayer} & \multicolumn{2}{c}{wire on
Pt} & \multicolumn{2}{c}{free wire}\\
& LMTO & FLAPW & LMTO & FLAPW & LMTO & FLAPW & LMTO & FLAPW & LMTO &
FLAPW \\ \hline 
$\langle\sigma_z\rangle$ direct& 1.83& 1.83& 2.03 & 2.00 & 2.02 & 2.12 &
2.21 & 2.06 & 2.22 & 2.16  \\ 
$\langle\sigma_z\rangle$ d only & 1.92 & 1.87 & 2.05 & 2.02 &
2.03 & 2.12 & 2.17 & 2.02 & 2.14 & 2.11 \\ 
$\langle\sigma_z\rangle$ from eq. (2)& 1.79 & 1.88 & 1.96 & 2.12 &
2.03 & 2.18 & 2.09 & 2.13 & 2.14  & 2.36 \\ 
$\langle\sigma_z\rangle$ from eq. (2), $\langle T_z\rangle=0$ & 1.80& 1.86& 
1.50 & 1.77 & 1.58 & 1.62 & 1.76 & 1.94 & 2.28 & 2.26 \\ 
$\langle T_z\rangle$ & 0.001& -0.003& -0.066 & -0.050& -0.064 &
-0.079& -0.048 & -0.027 & 0.021 & -0.014 \\ 
$\langle l_z\rangle$ direct & 0.15 & 0.15 & 0.13 &
0.13 & 0.18 & 0.19 & 0.14 & 0.14 & 0.71 & 0.40 \\ 
$\langle l_z\rangle$ from eq. (1) & 0.15 & 0.18  & 0.12 &
0.15 & 0.17 & 0.19 & 0.14 & 0.16 & 0.70 & 0.59 \\  
\hline
$\langle l_z\rangle$ direct& & 0.40& & 0.43 & & 0.86 & &0.92 & & 2.31\\
$\langle l_z\rangle$ from eq (1) & & 0.41& & 0.42 & & 0.86 & & 0.92 & & 2.17\\
\end{tabular}
\end{ruledtabular}
\caption{The spin and orbital moments (in Bohr magnetons) as calculated 
from the LSDA without the orbital-polarization term (upper part) 
and with the orbital-polarization term (lower part). For the meaning
of the various quantities, see text. The upper part gives also the 
$\langle T_z\rangle$ term appearing in the spin sum rule.}
\end{table*}
It should be noted again that thereby we use the same nearest-neighbor
distance between the atoms, i.e., the one of fcc Pt, for all
structures. As explained in Ref. \onlinecite{Ederer02}, the spin and
orbital moments derived experimentally via the sum rules correspond
essentially to the respective part of the valence wavefunctions with
3d angular momentum, if the ``background'' due to all additional
contributions to the experimental absorption spectra is subtracted
appropriately. We therefore compare in Table 1 the moments as
calculated directly from the 3d spin and orbital densities with those
obtained from the sum rules when taking into account for the
absorption spectra and for the $\langle T_z\rangle $ term also only
that part of the valence wave functions which has 3d character.  For
comparison, we give also the values of directly calculated moments
including all angular momentum contributions.
\par
For all the structures there is only a slight difference between the 
directly calculated moments with only d contributions and with all
contributions. The spin moments increase only moderately with decreasing
dimensionality. For the wire on Pt our directly calculated FLAPW spin
value of $2.06\>\mu_{\text{B}}$ agrees well with the one of Ref. 
\onlinecite{Buhlmayer?}.  There is also a satisfactory agreement between 
the directly calculated spin moments and those obtained from the 
spin sum rule when including the $\langle T_z\rangle$ term. 
The $\langle T_z\rangle$ contribution is negligible for hcp Co. 
In contrast, for the free and the Pt supported monolayers, the 
values of $\langle\sigma_z\rangle$ obtained from the sum rule when
neglecting $\langle T_z\rangle$ are between 20\% and 30\% smaller than
the directly calculated values. Astonishingly enough, when going to even
stronger reduced dimensionality, i.e., for the Pt supported wire and
even more for the free wire, the $\langle T_z\rangle$ term again 
appears to be of minor importance. To figure out the reason for this 
startling result, Fig. 2 shows the energy distribution\cite{Wu94/3} for
$\langle T_z\rangle$  for all the structures, i.e., the integral over the
energy resolved contribution to $\langle T_z\rangle$ up to an 
energy $E$ (for the integrals up to the Fermi level, $E=E_{\text{F}}$, 
$\langle T_z\rangle$ equals the quantity in eq. (2) ). 
\begin{figure}
\includegraphics[width=.45\textwidth]{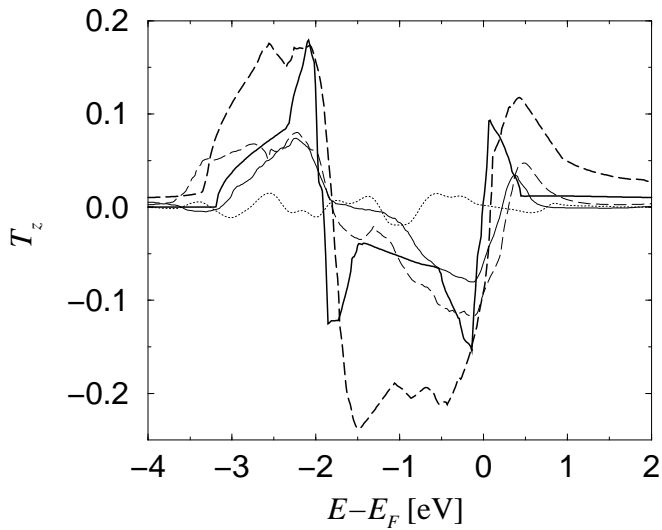}
\caption{The energy distribution of $\langle T_z\rangle$ for 
hcp Co (thin dotted line), a Co monolayer on Pt (thin dashed
line), a free standing Co layer (thick dashed line), a monatomic
Co wire on Pt (thin solid line) and a free standing monatomic Co wire 
(thick solid line).}
\end{figure}
It becomes obvious that for the free standing monolayer and for the
free standing wire, $\langle T_z\rangle$ depends drastically on $E$
resulting in sharp structures in the energy distribution due to the
two- or one-dimensional character of the systems. Going to the Pt
supported monolayer and wire, the energy dependence of $\langle
T_z\rangle$ is less pronounced and the sharp structures are smeared
out, indicating that these systems are not really two- or
one-dimensional due to the presence of the substrate. However,
$\langle T_z\rangle$ obtained from the integral up to $E_{\text{F}}$ is
very small for the free standing Co wire but larger for the other
systems (except for hcp Co).  This clearly demonstrates that the
expected rule of thumb --- a larger $\langle T_z\rangle$ for the
smaller dimensionality --- does not necessarily hold. We have
performed the same calculation for a free standing Ni wire.  The
energy distribution of $\langle T_z\rangle$ is very similar but the
Fermi level is shifted and $\langle T_z\rangle$ is consequently much
larger than for the Co wire. This again demonstrates that the symmetry
arguments alone are not able to estimate the size of $\langle
T_z\rangle$.
\par
The orbital moments obtained by the LSDA calculation with SO coupling
are very similar (about $0.15\>\mu_{\text{B}}$) for hcp Co and for the
Pt supported Co monolayer and wire, while the values are slightly
larger (about $0.18\>\mu_{\text{B}}$) for the free monolayer and
considerably larger for the free wire. Thereby, there is a good
agreement between the directly calculated orbital moments and those
obtained from the calculated spectra via the sum rules. The rather
small values for the Pt supported monolayer ($0.13\>\mu_{\text{B}}$) and
the Pt supported wire ($0.14\>\mu_{\text{B}}$) are in conflict with the
large corresponding experimental values\cite{Gambardella02}
($0.31\mu_{\text{B}}$ and $0.68\mu_{\text{B}}$). As discussed in
Ref. \onlinecite{Zhou99}, the reason is presumably that LSDA + SO
coupling does not appropriately account for orbital correlations. We
therefore have redone all the calculations by including in addition
the orbital-polarization term \cite{Eriksson90}.  Table 1, lower part,
gives the so obtained results for the orbital moments (the spin
moments are nearly unaffected by this additional term). It should be
noted that the calculations for hcp Co have been performed for the
nearest neighbor distance of fcc Pt which explains the large moment of
$0.4\>\mu_{\text{B}}$. For the Co monolayer on Pt (Co wire on Pt) our
value $0.43\>\mu_{\text{B}}$ ($0.92\>\mu_{\text{B}}$) is even larger than
the experimental value \cite{Gambardella02} $0.31\>\mu_{\text{B}}$
($0.68\>\mu_{\text{B}}$) which may be in part due to the fact that our
calculations did not take into account any structural relaxations at
the surface. For the free standing wire, the orbital moment appears to
be extraordinarily high ($2.3\>\mu_{\text{B}}$). Obviously the
experimental trend --- increase of the orbital moment with decreasing
dimensionality --- is well reproduced when taking into account the
orbital-polarization term. The increasing orbital moments result in an
increasing asymmetry between the calculated $L_2$ and $L_3$ XMCD
spectra (Fig. 3), in qualitative agreement with the respective
experimental spectra (Fig. 2 of Ref.  \onlinecite{Gambardella02}).
\begin{figure}
\includegraphics[width=.45\textwidth]{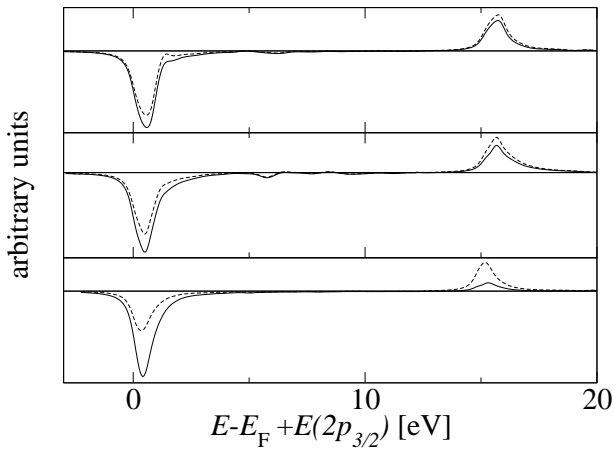}
\caption{The calculated XMCD spectra for hcp Co (upper part), 
Co monolayer on Pt (middle part) and Pt supported Co wire (lower
part). The dashed (solid) lines show the result from the LSDA without
(with) the orbital-polarization term.}
\end{figure}
\par
Both, for the LSDA calculations with and without the
orbital-polarization term, the directly calculated orbital moments
agree very well with those obtained from the sum rule. Therefore, for
Co in structures of various dimensionality the spin and orbital
moments probably may be obtained with high accuracy from the sum rules
(if including the $\langle T_z \rangle$ term in the spin sum rule).
\par
To conclude, we have shown that the orbital correlations have
to be taken into account explicitly in order to reproduce the experimentally
observed trend to higher orbital moments and larger asymmetries
between the L$_3$ and L$_2$ XMCD spectra in Co systems with reduced
dimensionality, and we are about to do this also for 
a systematic study of the magnetic anisotropy. In addition, we calculate
{\it ab initio} the further parameters appearing in a model Hamiltonian for 
the monatomic wires, i.e., exchange interactions from the 
adiabatic spin wave spectra \cite{Grotheer01}, and the Gilbert
damping factor \cite{Gilbert55}. The final objective is to 
come to a comprehensive description of the thermodynamic properties (see
Ref. \onlinecite{comment1}) and of the damped nonlinear
excitations (along the lines of Ref. \onlinecite{Holyst86}) of monatomic
magnetic linear chains. \par
The work at Brookhaven is supported by U.S. Department of Energy under Contract
No. DE-AC02-98CH10886.

\bibliography{wires.bbl}

\begin{thebibliography}{28}
\bibitem[{com()}]{comment1}
\bibinfo{note}{In this model the size of the superparamagnetic ``particles'' is
  estimated from the time-dependent kink density of the one-dimensional
  nearest-neighbor-exchange model in the infinite anisotropy (i.e., Ising)
  limit. Then these particles are considered as decoupled from each other but
  exposed to an external magnetic field and an anisotropy field, and the
  thermodynamics is described by classical Boltzmann statistics. It would be
  highly desirable to have a more rigorous treatment by calculating the
  dynamical structure factor for a one-dimensional chain with finite magnetic
  anisotropy and external magnetic field.}
\end{thebibliography}

\end{document}